\documentstyle[preprint,floats,pra,aps,psfig]{revtex}
\begin{document} \draft \textheight=9in \textwidth=6.5in

%----------------------------------------------------------------------

\title{\LARGE \bf Unitary reduction of the Liouville equation relative to a
two-level atom coupled to a bimodal lossy cavity}

\author{A. Napoli$^{1}$, Xiang-ming Hu$^{1,2}$, A. Messina$^{1}$}
\address{1)  INFM, MIUR and Dipartimento di Scienze Fisiche ed Astronomiche, via Archirafi 36, 90123 Palermo, Italy, Tel/Fax: +39 91 6234243; E-mail: messina@fisica.unipa.it\\
2)Department of Physics, Huazhong Normal University, Wuhan 430079, China }

\maketitle

\pagebreak

\begin{abstract}
The Liouville equation of a two-level atom coupled to a degenerate
bimodal lossy cavity is unitarily and exactly reduced to two
uncoupled Liouville equations. The first one describes a
dissipative Jaynes-Cummings model and the other one a damped
harmonic oscillator. Advantages related to the reduction method
are discussed.
\end{abstract}

\medskip

\pagebreak
\section{Introduction}
Over the last thirty years many theoretical studies have been done
in order to understand the dynamical nonlinear behavior of an atom
in a high-Q cavity. The interest toward this research area has
been mainly spurred by the huge number of experiments revealing
interesting features of the quantum radiation-matter coupling
\cite{Haroche}. Many theoretical and experimental activities
aimed, in particular, at understanding simple non trivial models
of quantum optics involving a single few-level atom and one or
more near resonant modes of the quantized electromagnetic field of
a cavity. The prototype of such models was that proposed by Jaynes
and Cummings (JCM) in 1963 \cite{JC} and describes a two-level
atom resonantly interacting with a single-mode field. The success
of this exactly solvable JCM has stimulated an intense research
devoted at highlighting and generalizing the original idea and
physical scenario \cite{SK}.

In this paper we concentrate our attention on the coupling between
two degenerate mode and an effective two-level atom. The dynamics
of this extended JCM model, under the hypothesis of ideal cavity,
has been recently studied \cite{Benivegna} bringing into light the
existence of purely and attractive quantum effects. It should be
noticed, however, that in realistic situations, cavities currently
used possess a finite quality factor $Q$ so that it is of interest
to know whether and how the predictions of the aforementioned
theory are affected by the relaxation of photons in the cavity. In
order to do this it is very usual to suppose the system coupled to
a reservoir represented as a bath of harmonic oscillators,
describing its dynamics by means of a master equation for its
density operator.

In this paper  we will show that the resolution of this problem
may be exactly traced back to that of two simpler decoupled
physical fictitious subsystems. We will indeed demonstrate that
the master equation describing the dynamics of a two-level atom
interacting with a degenerate bimodal lossy cavity is unitarily
equivalent to a system of two uncoupled and solvable Liouville
equations. The first one governs the dynamical behavior of a
dissipative one-mode JC model whereas the other one characterizes
the dynamics of a damped harmonic oscillator. As we shall see,
this kind of reduction of the original mathematical problem
provides a systematic and simple way to express the mean value of
a generic operator of the tripartite system under scrutiny, in
terms of expectation values relative to physical quantities
formally pertaining to the two fictitious subsystems.

\section{The model and its dynamics}
The model we consider consists of two electromagnetic modes of a
cavity interacting with an effective two-level atom of transition
frequency $\omega_0$.
The two independent cavity modes possess the same frequency
$\omega\sim \omega_0$ but they differ by polarization or direction of
propagation.
The effective two-mode JC Hamiltonian model describing,
in the Rotating Wave Approximation (RWA), such a tripartite system
can be cast in the following
form $(\hbar=1)$:
\begin{equation}
H=\sum_{\mu =1}^{2} \omega a_{\mu }^{\dagger }a_{\mu }+ \omega
_{0}S_{z}+\sum_{i=1}^{2} g_{\mu }\left( a_{\mu
}^{\dagger }S_{-}+a_{\mu }S_{+}\right) ,  \label{H}
\end{equation}
In equation\ (\ref{H}) the atomic degrees of freedom are, as usual,
represented by Pauli pseudo spin $ \frac{1}{2}$ operators while
$a_{\mu}$ $(a_{\mu}^{\dag})$ is the annihilation (creation) operator
of the $\mu$-th cavity mode. The two  real constants $g_1$ and $g_2$
measure the intensity of coupling between the atom and the cavity mode
denoted by 1 and 2 respectively.

Let's now observe that  the cavities at disposal of
experimentalists are characterized by a finite quality factor $Q$.
Realistic values of $Q$ have order of magnitude going from
$Q=10^{8}$, in correspondence to open Fabry-Perot resonator
\cite{QHaroche} to $Q=3\times 10^{10}$ for closed cavities
\cite{QWalther}. In order to take into account from the very
beginning the presence of cavity damping, we may suppose that our
system is coupled to a reservoir described as a bath of harmonic
oscillators \cite{Louisell}. We moreover  assume, as usual, that
the hypotheses under which the Born-Markov approximation may be
adopted, are satisfied \cite{Louisell}. Under these conditions,
the master equation for the density matrix $\rho $ of the combined
atom - two mode field system can be written in the form
\cite{Hu1,Hu2}
\begin{equation}
\frac{\partial \rho }{\partial t}=-i\left[ H,\,\,\rho \right]
+{\cal L}\rho ,  \label{m-e}
\end{equation}
where $H$ is the Hamiltonian given by eq.\ (\ref{H})
whereas, assuming zero temperature,
the losses of the cavity are represented by the superoperator ${\cal L}$
defined as
\begin{equation}
{\cal L}\rho =k \sum_{\mu =1}^{2} \left( 2a_{\mu }\rho a_{\mu
}^{\dagger }-a_{\mu }^{\dagger }a_{\mu }\rho -\rho a_{\mu }^{\dagger }a_{\mu
}\right)   \label{L}
\end{equation}
where $k$ is the damping constant. In eq.\ (\ref{L}) we have
realistically imposed that the damping constants relative to the
two cavity modes are equal.

\section{Unitary decoupling of the Liouville equation}
In order to solve equation \ (\ref{m-e}), we exploit the unitary
transformation realized by the operator $U$ defined as
\begin{equation}
U=e^{\gamma \left( a_{2}^{\dagger }a_{1}-a_{1}^{\dagger }a_{2}\right) }
\label{U}
\end{equation}
with $\gamma$ free real parameter to be appropriately fixed.

It is possible to demonstrate \cite{Benivegna} that the operator $U$
has the following properties:
\begin{eqnarray}
U^{\dagger }a_{1}U&=&\cos \gamma a_{1}+\sin \gamma a_{2} \nonumber
\\
U^{\dagger }a_{2}U&=&-\sin \gamma a_{1}+\cos \gamma a_{2}
\label{transf}
\end{eqnarray}

Taking into account equation\ (\ref{transf}), it is easy to prove that,
transforming the Hamiltonian $H$ by means the unitary operator $U$
defined by equation \ (\ref{U}), one obtains
\begin{eqnarray}
\tilde{H}&&\equiv U^{\dagger }HU=\tilde{H}_{1}+\tilde{H}_{2},
 \label{Htransf1} \\
\tilde{H}_{1}&&= \omega a_1^{\dagger }a_1+ \omega _{0}S_{z}+
g_{eff}\left( a_1^{\dagger }S_{-}+a_1S_{+} \right)
\label{Htransf2} \\
\tilde{H}_{2}&&=  \omega a_2^{\dagger }a_2,
\label{Htransf3}
\end{eqnarray}
where $g_{eff}=\sqrt{g_1^2+g_2^2}$, provided that
$\gamma=\tan^{-1}\left( g_{2}/g_{1}\right)$.

The transformed Hamiltonian $\tilde{H}$ describes two independent
subsystems in the sense that $\left[ \tilde{H}_1, \tilde{H}_2
\right]=0$. $\tilde{H}_1$ describes a simple (single mode)-(single
atom) system consisting of a fictitious radiation mode,
represented by the boson operators $a_1^{\dag}$, $a_1$, linearly
coupled to the old two-level atom with an effective coupling
constant $g_{eff}=\sqrt{g_1^2+g_2^2}$. $\tilde{H}_2$, on the other
hand, describes a new fictitious radiation mode, with creation and
annihilation operators given by $a_2^{\dag}$ and $a_2$
respectively, decoupled both from the atom and from the first
collective mode. In other words the canonical transformation of
$H$ accomplished by $U$ clearly brings into light the fact that
the two-level atom induces mode-mode coherence properties
responsible for the collective behavior of the field subsystem of
eq.\ (\ref{H}).

The circumstance that the Hamiltonian model\ (\ref{H}) is
unitarily equivalent to an one-mode JCM plus a free collective
mode, turns out to be the key for solving the master equation\
(\ref{m-e}). In order to better clarify this point let's start  by
observing that, if $\tilde{A}$ denotes the operator obtained
transforming by $U$ a generic operator $A$ of the system under
scrutiny, that is $\tilde{A}=U^{\dag}AU$, the master equation \
(\ref{m-e}) can be equivalently written in the form:
\begin{equation}
\frac{\partial \tilde{\rho} }{\partial t}=-i\left[ \tilde{H}
,\,\,\tilde{\rho} \right] +U^{\dag}{\cal L \rho}U
\label{m-e-transf}
\end{equation}
Exploiting eq.\ (\ref{L}) and\ (\ref{transf}), it is possible to
demonstrate that
\begin{equation}
U^{\dag}{\cal L}\rho U =
k \sum_{\mu =1}^{2} \left( 2U^{\dag}a_{\mu }U \tilde{\rho}
U^{\dag} a_{\mu
}^{\dagger }U-U^{\dag}a_{\mu }^{\dagger }a_{\mu }U \tilde{\rho}
-\tilde{\rho}U^{\dag} a_{\mu }^{\dagger }a_{\mu
}U\right) \equiv
{\cal
L}_1\tilde{\rho}+{\cal L}_2 \tilde{\rho} \label{Ltransf}
\end{equation}
where
\begin{equation}
{\cal L}_i \tilde{\rho}=
k \left( 2a_i \tilde{\rho} a_i^{\dagger }-a_i^{\dagger }a_i \tilde{\rho} -
\tilde{\rho} a_i^{\dagger }a_i \right) \;\;\;\;\; (i=1,2)
\label{L_i}
\end{equation}

Suppose now that the initial conditions imposed to the system are
such that the transformed density operator $\tilde{\rho}(0)$ can
be factorized into a product of two contributions,
$\tilde{\rho}_0^1$ relative to the atom and the collective mode 1
and $\tilde{\rho}_0^2$ describing the free fictitious mode 2. In
ref. \cite{Benivegna}, for example, the authors suppose that at
$t=0$ the system is prepared putting the atom in its ground state,
exciting one mode in a coherent state $\vert \alpha \rangle$ and
leaving the other one in its vacuum state. It is easy to
demonstrate that such an initial condition, in correspondence to
which the dynamics of the system manifests new and interesting
nonclassical features, is such that the condition
$\tilde{\rho}(0)=\tilde{\rho}_0^1\tilde{\rho}_0^2$ is satisfied.
Generally speaking, however, even if the initial condition imposed
to the system is such that the three subsystems, atom, mode 1 and
mode 2, are factorized, the transformed density operator
$\tilde{\rho}(0)$ at $t=0$ could not maintain the same
factorization property.

In the context of this paper we assume that
\begin{equation}
\tilde{\rho}(0)\equiv \tilde{\rho}_0^1\tilde{\rho}_0^2
\label{factorization}
\end{equation}

Our expectation, both on mathematical and physical grounds, is that the solution of eq.\ (\ref{m-e-transf}) satisfying such a factorized initial condition exists and is unique.

In a lossless situation eqs.\ (\ref{Htransf1})-\ (\ref{Htransf3})
guarantee that the density matrix operator at a generic time
instant $t$ keeps its initial factorized form. We may wonder
whether the solution of equation\ (\ref{m-e-transf}) satisfying
the initial condition given by eq.\ (\ref{factorization}) still
possesses such a property. We may indeed show that this is the
case. To this end we state and prove the following
\textit{Theorem}: The unique solution of equation\
(\ref{m-e-transf}) satisfying the initial condition\
(\ref{factorization}) may be given in the form
$\tilde{\rho}(t)=\tilde{\rho}_1(t)\tilde{\rho}_2(t)$ where
$\tilde{\rho}_1(t)$ and $\tilde{\rho}_2(t)$ are solutions of the
two following Cauchy problems:
\begin{eqnarray}
\frac{\partial \tilde{\rho}_{1}}{\partial t} &=&-i\left[
\tilde{H}_{1},\,\,\tilde{\rho}_{1}\right] +{\cal L}_{1}\tilde{\rho}_{1},
 \;\;\;\;\;
\tilde{\rho}_1(0)=\tilde{\rho}_0^1
\label{m-e-1} \\
\frac{\partial \tilde{\rho}_{2}}{\partial t} &=&-i\left[
\tilde{H}_{2},\,\,\tilde{\rho}_{2}\right] +{\cal L}_{2}\tilde{\rho}_{2}
\;\;\;\;\;
\tilde{\rho}_2(0)=\tilde{\rho}_0^2
\label{m-e-2}
\end{eqnarray}

\textit{Proof}. Looking for solutions of equation\
(\ref{m-e-transf}) in the factorized form
$\tilde{\rho}(t)=\tilde{\rho}_1(t)\tilde{\rho}_2(t)$ yields

\begin{equation}
[\frac{\partial \tilde{\rho_1} }{\partial t}+i \left[ \tilde{H}_1
,\,\,\tilde{\rho}_1 \right]-{\cal L}_1 \tilde{\rho}_1]\tilde{\rho}_2+
\tilde{\rho}_1[\frac{\partial \tilde{\rho_2} }{\partial t}+
i \left[ \tilde{H}_2,\,\,\tilde{\rho}_2 \right]-{\cal L}_2 \tilde{\rho}_2]
=0
\label{2m-e}
\end{equation}

As first step we now prove that eq.\ (\ref{2m-e}) necessarily implies
\begin{eqnarray}
\frac{\partial \tilde{\rho}_{1}}{\partial t} +i\left[
\tilde{H}_{1},\,\,\tilde{\rho}_{1}\right] -{\cal L}_{1}\tilde{\rho}_{1}
=\lambda_1\tilde{\rho}_{1},
\label{l1} \\
\frac{\partial \tilde{\rho}_{2}}{\partial t} +i\left[
\tilde{H}_{2},\,\,\tilde{\rho}_{2}\right] -{\cal L}_{2}\tilde{\rho}_{2}=
\lambda_2 \tilde{\rho}_{2}
\label{l2}
\end{eqnarray}
with $\lambda_1=-\lambda_2$.

To this aim, for the sake of simplicity,
we rewrite equation\ (\ref{2m-e}) in the following form:
\begin{equation}
A_1B_2+A_2B_1=0
\label{AB}
\end{equation}
where the operators $A_i\equiv \frac{\partial
\tilde{\rho}_{i}}{\partial t} +i\left[
\tilde{H}_{i},\,\,\tilde{\rho}_{i}\right] -{\cal L}_{i}\tilde{\rho}_{i}$
and $B_i\equiv \tilde{\rho}_i$ $(i=1,2)$ are defined in the Hilbert
space of the system $i$.
We now prove that, for each fixed $i$ $(i=1,2)$, it must be:
\begin{equation}
A_i=\lambda_iB_i
\label{A=lB}
\end{equation}
To this end let's consider a generic basis $\{ \vert n_1 \rangle
\otimes \vert n_2 \rangle \}$ of the Hilbert space of the total
system obtained as a tensorial product of two basis, $\{ \vert n_1
\rangle\}$ and $\{ \vert n_2 \rangle \}$ relative to the Hilbert
space of the system 1 and 2 respectively. The action of both
members of equation\ (\ref{AB}) on a generic state vector $ \vert
n_1 \rangle \otimes \vert n_2 \rangle $ of the prefixed basis in
the Hilbert space of the total system (1)+(2) may be formally put
as follows:
\begin{eqnarray}
&&(A_1B_2)(\vert n_1 \rangle \otimes \vert n_2 \rangle)+
(A_2B_1)(\vert n_1 \rangle \otimes \vert n_2 \rangle)=
 \nonumber \\
&&=(A_1 \vert n_1 \rangle )(B_2 \vert n_2 \rangle )+
(A_2 \vert n_2 \rangle )(B_1 \vert n_1 \rangle)=0 \label{ABn}
\end{eqnarray}

Suppose now $A_1\not= \lambda_1B_1$, in contradiction with eq.\
(\ref{A=lB}). Then there exists at least one vector, say $\vert
\bar{n}_1 \rangle$ such that
\begin{equation}
A_1 \vert \bar{n}_1 \rangle \not= \lambda_1 \vert \bar{n}_1 \rangle
\end{equation}
Accordingly equation\ (\ref{ABn}) may be cast in the form
\begin{equation}
(A_1 \vert \bar{n}_1 \rangle )(B_2 \vert n_2 \rangle )+
(A_2 \vert n_2 \rangle )(B_1 \vert \bar{n}_1 \rangle)=0 \label{ABn1}
\end{equation}
whatever $\vert n_2 \rangle$ is.
This implies that for any $\vert p_2 \rangle$
\begin{equation}
\langle p_2 \vert B_2 \vert n_2 \rangle=
\langle p_2 \vert A_2 \vert n_2 \rangle=0
\end{equation}
which in turn leads to the conclusion $A_2=B_2=0$ evidently false
since $Tr\{ B_2\}=1$ at any $t$. Thus $A_1= \lambda_1B_1$ and,
repeating the same argument after exchanging 1 with 2, $A_2=
\lambda_2B_2$. It is moreover easy to verify with the help of eq.\
(\ref{ABn}), that necessarily $\lambda_1=-\lambda_2\equiv\lambda$.
This concludes the first step of our proof.

The second and last step in the proof of our initial statement consists
in proving that the only physically admissible solution is that
correspondent to $\lambda=0$.
Let's indeed consider the general equation
\begin{equation}
\frac{\partial \rho}{\partial t} +i\left[
H,\,\,\rho\right] -{\cal L}\rho=
\lambda \rho
\end{equation}

Taking the trace of both members of this equation we
obtain:
\begin{equation}
\frac{\partial  }{\partial t}Tr \{\rho\} -Tr\{{\cal L}\rho\}=
\lambda Tr \{ \rho\}
\label{Lro}
\end{equation}
since $Tr \{\left[H,\,\,\rho\right]\}=0$. Exploiting, moreover,
elementary algebraic properties of the trace and taking into
account the explicit form of the superoperator ${\cal L}$, given
by eq.\ (\ref{L}) ignoring the subscript $i$, it is possible to
demonstrate that also $Tr\{{\cal L}\rho\}$ is equal to zero. In
view of these considerations eq.\ (\ref{Lro}) becomes:
\begin{equation}
\frac{\partial  }{\partial t}Tr \{\rho\} =
\lambda Tr \{ \rho\}
\label{Lror}
\end{equation}

We thus may state that, being $Tr \{ \rho\}=1$, eq.\ (\ref{Lror})
necessarily implies $\lambda =0$.

Before concluding this section we wish to underline that
the unitary reduction of the master equation \ (\ref{m-e})
into a system of two uncoupled Liouville equations
(see eqs. \ (\ref{m-e-1}) and \ (\ref{m-e-2})),
paves the way to the exact treatment of the dynamics of
the physical system under scrutiny.

Let's indeed consider a generic operator
$O=f(a_1,a_1^{\dag},a_2,a_2^{\dag},\hat{\textbf{S}})$ of the
system. In order to calculate the mean value $\langle O \rangle$
of this operator at a generic time instant $t$, we observe that
\begin{equation}
\langle O \rangle=Tr \{ \rho O \}=
Tr \{ (U^{\dag}\rho U)(U^{\dag} O U) \}
\label{O}
\end{equation}
But
\begin{equation}
U^{\dag}\rho U=\tilde{\rho}\equiv \tilde{\rho}_1 \tilde{\rho}_2
\end{equation}
and
\begin{equation}
U^{\dag} O U \equiv U^{\dag}f(a_1,a_1^{\dag},a_2,a_2^{\dag},
\hat{\textbf{S}})U
=f(U^{\dag}a_1U,U^{\dag}a_1^{\dag}U,U^{\dag}a_2U,U^{\dag}a_2^{\dag}U,
\hat{\textbf{S}})
=\tilde{f}(a_1,a_1^{\dag},a_2,a_2^{\dag},\hat{\textbf{S}})
\end{equation}
where $\tilde{f}$ is a new function of the operators $a_1$,
$a_1^{\dag}$, $a_2$, $a_2^{\dag}$ and $\hat{S}$ now related to the
two fictitious subsystems 1 (atom - mode 1) and 2 (mode 2)
obtained after the transformation defined by the unitary operator
$U$.

Let's now observe that in many physical problems of interest,
$\tilde{f}$ has the form of a linear superposition of products of
operators $a_i$, $a_i^{\dag}$, $(i=1,2)$ and $\hat{\textbf{S}}$
that is
\begin{equation}
\tilde{f}(a_i,a_i^{\dag},\hat{S})=\sum_{k}\sum_{i}{f_k
g^i_k(a_i,a_i^{\dag})g_k^3(\hat{S})}
\label{tildef}
\end{equation}
where $g^i$ $(i=1,2)$ and $g^3$ are prefixed functions. Replacing
$U^{\dag}OU$ appearing in equation\ (\ref{O}) by $\tilde{f}$ as
expressed by eq.\ (\ref{tildef}) immediately yields
\begin{equation}
\langle O \rangle =\sum_{k}{f_k Tr
\{\tilde{\rho}_1g^1_k(a_1,a_1^{\dag})g_k^3(\hat{S})\}
Tr\{\tilde{\rho}_2g_k^2(a_2,a_2^{\dag})\}} \label{end}
\end{equation}
Eq.\ (\ref{end}) says that the mean value of the operator $O$ can
be expressed as combination of mean values of operators relative
to the two fictitious subsystems 1 and 2 respectively. In any
case, whatever the form of the function $f$ is, exploiting the
already known solutions $\tilde{\rho}_1$ and $\tilde{\rho}_2$,
describing the dissipative JC model and the damped harmonic
oscillator respectively, eq.\ (\ref{end}) provides a simpler way
to evaluate the mean value of a generic operator $O$ of our
original physical system.

\section{Conclusive remarks}
In this paper we have considered an effective two-level atom
resonantly coupled to a degenerate bimodal lossy cavity. Our main
and novel result is that, under appropriate but general enough
initial conditions, the Liouville equation of our tripartite
system can be unitarily and exactly reduced to a pair of uncoupled
and simpler master equations describing a damped harmonic
oscillator and a dissipative one-mode JC model.

The success of our procedure is intimately related to the
knowledge of a unitary operator accomplishing the decoupling of
one bosonic degree of freedom from the other dynamical variables.

We emphasize that since both these new  fictitious problems may be
analytically solved, our recipe provides the key for an exact
analytical treatment of the original Liouville equation. A
complete investigation of this dynamical problem is behind the
scope of this paper and will be presented elsewhere.

An effective way to simplify the evaluation of expectation values
of physical quantities of interest, with the help of our method,
has also been presented and briefly discussed.

We conclude underlining that the reduction method reported in this
paper might also be useful to treat master equations relative to
other physical systems. As an example, it is easy to verify that
our procedure is successfully applicable to the hamiltonian model\
(\ref{H}) revised including the counter rotating terms.

\pagebreak

{\LARGE \bf Acknowledgments} One of the authors (A.N.)
acknowledges financial support from Finanziamento Progetto Giovani
Ricercatori anno 1999, Comitato 02. This work was partially
supported by National Natural Science Foundation of China under
Grant No 10074018 and supported by Foundation for University Key
Teacher by Ministry of Education (GG-140-10511-1004). X.M.H wishes
to thank Professor J.-S. Peng for discussions on this subject.

%%%%%%%%%%%%%%%%%%%%%%% REFERENCES %%%%%%%%%%%%%%%%%%%%%%%%%%%%%%%

\end{document}